\documentclass[sigconf]{acmart}

\AtBeginDocument{%
  }

\setcopyright{acmcopyright}
\copyrightyear{2018}
\acmYear{2018}
\acmDOI{XXXXXXX.XXXXXXX}

\acmConference[Conference acronym 'XX]{Make sure to enter the correct
  conference title from your rights confirmation email}{June 03--05,
  2018}{Woodstock, NY}
\acmPrice{15.00}
\acmISBN{978-1-4503-XXXX-X/18/06}

\usepackage{multirow}
\usepackage{siunitx}

\begin{document}

\title{Experiential Benefits of Interactive Conflict Negotiation Practices in Computer-Supported Shift Planning}


\author{Alarith Uhde}
\email{alarith.uhde@uni-siegen.de}
\orcid{0000-0003-3877-5453}
\affiliation{%
  \institution{University of Siegen}
  \streetaddress{Kohlbettstraße 15}
  \postcode{57072}
  \city{Siegen}
  \country{Germany}
}
\author{Matthias Laschke}
\email{matthias.laschke@uni-siegen.de}
\orcid{0000-0003-2714-4293}
\affiliation{%
  \institution{University of Siegen}
  \streetaddress{Kohlbettstraße 15}
  \postcode{57072}
  \city{Siegen}
  \country{Germany}
}
\author{Marc Hassenzahl}
\email{marc.hassenzahl@uni-siegen.de}
\orcid{0000-0001-9798-1762}
\affiliation{%
  \institution{University of Siegen}
  \streetaddress{Kohlbettstraße 15}
  \postcode{57072}
  \city{Siegen}
  \country{Germany}
}


\begin{abstract}

  Shift planning plays a key role for the health and well-being of healthcare
  workers. It determines when they work and when they can take time off to
  recover or engage in social activities. Current computer-support in shift
  planning is typically designed from a managerial perspective and focuses on
  process efficiency, with the long-term goal of full automation. This implies
  automatic resolutions of emotionally charged scheduling conflicts. In the
  present study, we measured the effects of such a fully automated process on
  workers' well-being, fairness, and team spirit, and compared them with a more
  interactive process that directly involves workers in the decision-making. In
  our experimental online study (n = 94), we found positive effects of the more
  interactive process on all measures. Our findings indicate that full
  automation may not be desirable from the worker perspective. We close with
  concrete suggestions to design more worker-centered, hybrid shift planning
  systems by optimizing worker control, considering the worker experience, and
  embedding shift planning in the broader work context.

\end{abstract}



\maketitle

\section{Introduction}

Shift planning is an essential work organization process in many healthcare
institutions. It sets the general frame for all care activities by defining who
works on which days and who can stand in if someone calls in sick. Shift
planning also has implications for each healthcare worker's private life,
health, and well-being, because it determines when they can sleep or participate
in social events outside of work~\citep{Costa2010, Fenwick2001}. Certain
planning decisions around unpopular shifts can affect fairness and the
interpersonal relationships in the team (i.e., the ``team
spirit'')~\citep{Almost2010, Schlicker2021a, Uhde2020a}. In sum, shift planning
is an important work organization process in healthcare that has considerable
consequences for the private and work lives of healthcare workers.

A perfect shift planning process would manage to integrate the personal needs of
all workers with the requirements of the healthcare institution. Of course, that
is not always possible. Sometimes many workers want to spend time with their
friends or families during certain shifts (e.g., around public holidays such as
Christmas), and the remaining workers cannot cover the demand to assure patient
safety. In such cases, all possible solutions imply that some additional workers
will have to be assigned to work and cancel their private plans. These difficult
decisions with personal consequences for individual healthcare workers are an
inextricable part of shift planning.

Current computer-supported shift planning systems can automatically identify
such situations in which conflicts arise between workers' requests for free
shifts on the one hand, and requirements of the healthcare institution on the
other. They can also make reasonable decisions to automatically resolve
them~\citep{Burke2004}. These computer-supported systems rely on algorithmic
problem-solving techniques and use formalized data about worker requests, shift
lengths, legal regulations, worker qualifications, and many other parameters
considered relevant~\citep{VandenBergh2012}. The algorithm identifies a planning
conflict if there is no possible solution that satisfies all legal regulations
and also all workers' requests in the formalized representation. To solve this
problem, a typical algorithm removes some ``less important'' restrictions such
as a worker's request to find a valid solution~\citep{Burke2004}. Such
computational approaches allow for efficient, fully automated shift planning
without the need for further human interventions.

However, in practice, some planning conflicts can have solutions that are
difficult to detect automatically, because the algorithm does not have all
relevant context information. For example, two healthcare workers could have
a request for a free shift on a particular Friday evening, but one of them needs
to be assigned to work. The first worker has planned to meet with an old friend,
the second worker wants to go to a concert. The concert has a fixed time, but
one way to resolve this planning conflict is to call the friend and try to
reschedule their meeting to the Thursday or Saturday. This solution would still
enable both workers to integrate their private needs with the schedule.
However, the process requires some manual communication with the friend. In
another example, one worker may want to go to a party that starts at 6pm, but
their colleague cannot take over the shift before 7pm. If it would still be
acceptable for the first worker to arrive at the party one or two hours late,
the colleagues can still make it work in practice. But this solution requires
some communication with the worker to see whether they are flexible. Otherwise,
an automatic system would simply identify a worker shortage between 6pm and 7pm.
Some other context information is inaccessible to the algorithm because of
privacy regulations (e.g., personal relationships between workers), or because
entering all details that might be important for each request requires a lot of
(often unnecessary) effort. In other words, fully automated conflict resolution
techniques can only find solutions based on the available data, and they oversee
some solutions that require more context information and flexibility of the
involved people.

Each individual request for free time can represent a subjectively important
private appointment for a healthcare worker. In their everyday work lives, they
often have to work at unusual hours or on weekends, and organizing a fulfilling
social life under these circumstances is already difficult~\citep{Costa2016,
Fenwick2001}. Thus, each granted request can contribute to a better work-life
balance for an individual healthcare worker, and shift planning systems should
be designed to support workers with their private plans as much as possible.
From our perspective, the primary goal of conflict resolution between workers'
requests should be to find a solution that works for everyone involved, even if
it may be slightly less efficient than a fully automated alternative.

Because requests for free time are usually important for the healthcare workers,
negotiations to resolve planning conflicts can become emotional. On the one
hand, depending on individual negotiation styles~\citep{Friedman2000} and
contextual factors such as time pressure~\citep{Brown2011}, conflict
negotiations may lead to negative emotional experiences. On the other hand, they
also represent an occasion for pro-social interactions between workers, which
can create positive experiences~\citep{Uhde2021a}. For example, healthcare
workers can help their colleagues by being flexible with their own requests
(e.g., rescheduling it or going later to a party), or they may receive help
themselves. Involvement in the decision-making process of conflicts that affect
workers' lives is essential for their fairness experiences~\citep{Uhde2020a},
which can contribute to their satisfaction with the work schedule and work-life
balance~\citep{Nelson2010}. In addition, a positive team spirit can also support
a cooperative attitude during conflict negotiations~\citep{Uhde2020a}. All in
all, conflict negotiations can have experiential consequences for workers, and
we think that they should be designed to support more positive experiences.

In this paper, we investigated how interactive conflict negotiations can be
integrated in computer-supported shift planning, so that they add to a positive
experience for healthcare workers. Our long-term goal is not to replace
automatic shift planning entirely and give up its benefits of efficiency.
Instead, we want to work more specifically towards improved worker experiences
of conflict resolutions, which we consider the most critical phases of shift
planning from the workers' perspective. This paper builds on several interactive
conflict negotiation practices developed by healthcare workers in previous work,
which helped them find solutions that integrate their own needs with the needs
of their colleagues and the shift plan~\citep{Uhde2021a}. Our main contribution
are the results from an experimental study in which we could confirm that such
interactive practices actually have positive consequences on workers'
well-being, fairness experiences, and their team spirit, compared with a fully
automated process.

\section{Related Work}

\subsection{Effects of Worker Involvement in Shift Planning}

Overall, shift work has negative effects on workers' health and well-being. It
affects their sleep~\citep{Brown2015}, psychological and physiological
health~\citep{Bohle1989, Knutsson1989, Costa2010}, and social
life~\citep{Arlinghaus2019, Barton1995, Loudoun2008, Walker1985}. Accordingly,
healthcare workers, who often work in shifts, have relatively low average
subjective well-being, compared with the general population~\citep{Oates2017,
Oates2018}. Problems related to shift work are a main reason why healthcare
workers consider leaving the profession~\citep{Chan2013, Flinkman2008} and they
are a reason for young people not to choose a career as a healthcare
worker~\citep{Hemsley1999}.

Some negative effects of shift work can be reduced through improvements of the
shift plans and the planning process. For example, the order of shifts (i.e.,
morning, evening, night) and the shift lengths can be adjusted for each worker
to support more healthy sleep patterns~\citep{Brown2015, Costa2010, DallOra2016,
Folkard2007}. More stable and predictable shift plans can also help workers
pursue a more fulfilling social life~\citep{Costa2006}. Finally, giving workers
more control of their shift plans can mitigate some negative impacts of shift
work on health and well-being~\citep{Fenwick2001, Garde2012, Kubo2013}.

Besides these effects on well-being, the shift planning process also affects the
group dynamics in the team. Critical planning decisions, for example about who
has to work despite having personal plans, have implications for healthcare
workers' fairness experiences and team spirit~\citep{Uhde2020a, Schlicker2021a}.
Involving workers in the decision-making process increases fairness and gives
room for negotiations between colleagues that can affect the team
spirit~\citep{Uhde2020a}. Moreover, the effects of worker involvement on team
spirit, well-being, and fairness can reinforce each other on the long run. For
example, a positive team spirit can reduce further work-life conflicts, which in
turn supports well-being~\citep{Pisarski2001, Pisarski2008}.

Despite these positive effects, worker involvement can also have negative
effects from an organizational perspective. Interactive shift planning that is
based on manual interactions (e.g., ``self-scheduling''~\citep{Miller1984}) can
be less efficient than automatic shift planning~\citep{Burke2004}. It can also
be impractical in larger teams, where more people need to interact with all of
their colleagues~\citep{Silvestro2000}. In some cases, shift planning based on
manual interactions can lead to invalid shift plans that, for example, do not
comply with legally required minimal staffing levels~\citep{Bailyn2007}.
Finally, automatic shift planning can lead to more consistent decisions across
different teams~\citep{Silvestro2000}.

\subsection{Computer-Supported Shift Planning and Worker Involvement}

These advantages of automation from an organizational perspective have motivated
a long tradition of research on computer-supported shift planning
(e.g.,~\citep{Miller1976a, Warner1976, Burke2004, VandenBergh2012}). Most of
these computer-supported systems are based on an automatic process with minimal
worker involvement~\citep{Burke2004, VandenBergh2012}. In some of these systems,
workers can submit requests for free shifts (e.g.,~\citep{Miller1976a,
Lin2015}), and optionally provide additional metadata with their requests, such
as a subjective importance rating~\citep{Constantino2011}. The planning
algorithm uses these data to make autonomous decisions.

Some more recent computer-supported approaches have focused more specifically on
worker involvement. But their effects on workers varied with the specific
implementations. One problem of these previous studies is that they report only
summative effects of the system as a whole and do not link the effects to
specific design decisions. For example, Rönnberg and
Larsson~\citep{Roennberg2010} allowed workers to submit their preferred, full
shift plans. Possible conflicts were resolved by the head nurse. This process
led to high quality shift plans compared with the previous, manual process
(i.e., legally compliant). But healthcare workers suspected that they might lose
influence about their work times. Nabe-Nielsen and
colleagues~\citep{NabeNielsen2012} implemented a similar process, but
additionally allowed healthcare workers to make changes to a preliminary version
of the group schedule. The healthcare workers experienced this shift planning
process as more flexible than two alternatives using fixed shift plans, but also
as less predictable.

Submitting full shift plans potentially maximizes control, but it adds work load
for the healthcare workers and thus causes new problems. For example, Bailyn and
colleagues~\citep{Bailyn2007} reported a case where an approach using full
worker-submitted shift plans has failed. Healthcare workers were supposed to
collaboratively create almost the entire shift plan by themselves. The head
nurse published a draft with some prefilled shifts (because a few workers had
fixed shift plans), and relied on the other healthcare workers to fill in the
remaining shifts and to create a functional shift plan. After a while, some
shifts were not covered appropriately anymore, and the workers stopped engaging
in collaborative shift planning and conflict resolution. The head nurse had to
intervene and eventually returned to their previous planning process. The extra
work, a lack of collaborative behavior, and possibly the large group size (over
70 registered nurses) may have contributed to these problems~\citep{Bailyn2007}.
Thus, despite short-term positive effects (e.g., higher subjective control of
the schedule, better quality of patient care), the shift plans became
inconsistent after a while, and the system was unsustainable on the long run.

Uhde and colleagues~\citep{Uhde2021a} followed a different approach and asked
workers to only plan for appointments that are (subjectively) important to them,
while leaving everything else to the computer to plan automatically. For the
less important shifts, healthcare workers could submit more abstract, long-term
preferences. For example, they could submit that they generally preferred to
work in the morning, or that they have a weekly recurring appointment (e.g.,
a sports club). Workers were supposed to resolve conflicts among themselves, and
they managed to find solutions that integrated everyone's needs in most cases.
In terms of actual involvement, some healthcare workers used the system more
actively and submitted many requests, while others hardly used it at all,
because they generally tended to make fewer private plans. One problem for the
workers in this study was that the shift plans were only released two weeks in
advance, which reduced predictability.

\subsection{Conflict Resolution and Worker Experience}

As indicated above, most of the previous work on computer-supported shift
planning excludes healthcare workers from the decision-making process. The few
studies that involve workers reported more general findings of the shift
planning approach as a whole. Here, we summarize previous work that is more
specifically focused on conflict resolution in (typically manual) healthcare
shift planning.

Several factors that relate to healthcare workers' experiences of conflict
negotiations have been studied as antecedents or process aspects, including
negotiation styles, team spirit, and contextual factors such as time pressure
and work load~\citep{Almost2010, Brown2011, Pisarski2001, Pisarski2008}.
Integrative negotiation styles can lead to positive
experiences~\citep{Friedman2000}, and are the most common style among healthcare
workers~\citep{Labrague2018}. In contrast, avoiding or ``dominating''
negotiation styles~\citep{Friedman2000} and the use of harsh
language~\citep{Rogers2011} can lead to negative experiences during conflict
negotiations.

Fairness of decision-making processes such as conflict negotiations also plays
an important role. It is associated with fewer interpersonal
conflicts~\citep{Oxenstierna2011}, higher satisfaction with the shift plan, and
a better subjective work-life balance~\citep{Nelson2010}. Fairness is often
studied in work settings where the decisions are made by
a supervisor~\citep{Schlicker2021a, Lee2015, Oetting2018}. But a few studies
have focused on decisions made by workers themselves~\citep{Uhde2020a,
Uhde2021a}. In healthcare shift planning more specifically, workers' subjective
fairness experiences are based on two different fairness
norms~\citep{Deutsch1975}. For generally desirable resources (e.g., free
weekends, holidays), healthcare workers' fairness experiences depend on the
equality norm, which means that they consider shift plans fairer if such
resources are distributed equally among workers~\citep{Uhde2020a}. However,
subjective fairness of specific conflict resolutions is based on the need norm,
meaning that the decision should be based on the individual needs of everyone
involved~\citep{Deutsch1975, Uhde2020a}. For example, if one worker is a single
parent and has to take care of their child, their need is typically considered
higher than the need of a colleague who wants to go to the cinema. One problem
with the need norm in automatic shift planning is that the needs cannot easily
be formalized and compared. Thus, conflict resolution based on direct worker
involvement has been suggested~\citep{Uhde2020a}.

One previous system has been presented that included an explicit face-to-face
negotiation process among workers in otherwise automated shift
planning~\citep{Uhde2021a}. However, healthcare workers did not use the system's
software-based process directly, and instead developed several interactive
conflict negotiation practices themselves. These interactive conflict
negotiation practices represent the concrete actions healthcare workers took to
negotiate shift planning conflicts among each other and with the shift planning
system. Healthcare workers first tried to reschedule their private appointments
if possible, to avoid a conflict altogether (e.g., meeting with a friend on
Thursday instead of Friday, if a co-worker already has a request on that
Friday). If that was not possible because of the nature of the request (e.g.,
a concert with a fixed time), the second step was to try and activate
alternative resources, such as colleagues from other wards or splitting the
shift. Third, some conflicts were expectable, for example around public
holidays, and the workers used informal team rules during group meetings to find
a solution. Finally, if none of these practices worked, conflicts were resolved
by the management.

In conflict negotiations based on interactions between healthcare workers, team
spirit becomes important, because it affects workers' openness to
negotiate~\citep{Uhde2020a}. A negative team spirit can lead to interpersonal
conflicts in the team~\citep{Almost2010}. It can also negatively affect conflict
management styles, and lead to conflict avoidance and a lack of consideration
for others' needs~\citep{Almost2010, Almost2016}.

\subsection{Summary and Research Questions}

Overall, shift work has negative effects on workers' well-being. These negative
effects can be mitigated in part by involving workers in decision-making
processes, particularly in conflict negotiations around their requests for free
time. During such conflict negotiations, fairness and a positive team spirit
become important, and they can be part of an overall positive experience. In
addition, successful negotiations can help workers better integrate their
private and work lives, with positive long-term effects on their health and
well-being.

Computer-supported shift planning has several advantages in terms of efficiency
and the quality of shift plans, but it often lacks worker involvement. Some
recent, hybrid systems have attempted to integrate the advantages of
computer-supported shift planning with selective worker inclusion. For example,
automatic shift planning can be used for all (subjectively) non-critical shifts,
and workers can resolve the remaining conflicts about their important requests
manually. Such a hybrid approach could lead to better overall shift planning
that is both efficient and has positive effects on workers. Our goal in this
study is to investigate the experiential consequences of concrete, interactive
conflict negotiation practices, that is, the specific activities workers can
take during conflict negotiation to find solutions that reconcile their private
needs with organizational requirements.

Based on previous literature, we assumed positive effects of worker inclusion on
subjective well-being, fairness, and team spirit. This led to our four
hypotheses:

\begin{description}
  \item [H1:] Interactive conflict negotiation practices lead to a more positive
    emotional experience for workers (i.e., affect balance) than fully automated
    conflict resolution.
  \item [H2:] Interactive conflict negotiation practices lead to more need
    fulfillment than fully automated conflict resolution.
  \item [H3:] Interactive conflict negotiation practices are perceived as fairer
    than fully automated conflict resolution.
  \item [H4:] Interactive conflict negotiation practices lead to a better
    long-term team spirit than fully automated conflict resolution.
\end{description}

\section{Method}

We used the experimental vignette methodology (EVM;~\citep{Aguinis2014}) to test
our hypotheses in an online study. EVM is an experimental method based on
immersive scenarios, typically presented as short texts. It allows researchers
to experimentally control situational factors and to vary certain aspects of
interest. In our specific case, we chose EVM because it allowed us to
experimentally test the effect of specific interactive conflict negotiation
practices, independent of their outcome in practice (i.e., whether a worker's
request was granted or not). Thus, EVM allowed us to provide enough context to
create and compare realistic scenarios while removing other factors that may
influence the participants' experiences in field studies. EVM has been widely
used in Human-Computer Interaction~\citep{Diefenbach2017c, Oetting2018,
Uhde2020a} and beyond~\citep{Atzmueller2010, Tversky1981}.

\subsection{Participants}

In total, 94 German-speaking healthcare workers completed our online study. We
recruited participants through snowball sampling, social media channels, and
indirectly by asking the healthcare managers of around fifty healthcare
institutions to forward our call. As compensation, participants could
participate in a raffle for a 30 € voucher, and we additionally donated 50 Cents
for each complete questionnaire to a non-profit organization to increase
motivation to participate.

The sample included 13 participants who identified as male and 81 who identified
as female. The average age was 39 years ($sd = 11.20$; $min = 21$; $max = 63$).
Participants had 16 years of work experience on average ($sd = 12.10$; $min
= 0.5$; $max = 46$). 62 participants worked as registered nurses, and another 17
had an additional role as ward leader. 6 participants worked as childcare
nurses, 3 as nurse assistants, 3 had other healthcare-related occupations
(intern, midwife, healthcare specialist), and 3 did not disclose their specific
occupation.

\subsection{Procedure}

The participants accessed the study through a link or a QR code provided during
the advertisement for the study. They landed on a welcome page that informed
them about the overall process of the study, the expected length (15 minutes),
and anonymous data analysis. On the next screen, participants were introduced to
the general scenario, which was based on a hybrid shift planning model inspired
by~\citep{Uhde2021a}. We asked them to imagine that a new scheduling system was
introduced in their team. They could access this new system through an app on
a tablet in their office. The scheduling system allowed each healthcare worker
to submit long-term preferences for early shifts, late shifts, or specific times
in a week (e.g., Wednesday from 16:00 to 18:00, so they could join their sports
club). We said that these preferences only needed to be entered once and were
valid until changed. In addition, the shift planning system allowed workers to
enter requests for specific free days or shifts.

On the following pages, we presented four vignettes of planning scenarios in
relation to such requests. In the ``conflict negotiation'' condition, we used
the four interactive conflict negotiation practices presented
in~\citep{Uhde2021a}. Healthcare workers could try to find a solution using
these practices, and used a fully automated process as a fallback. In the
``automatic'' condition, the process consisted only of the fully automated
processes, as is common in many shift planning systems (e.g.,~\citep{Burke2004,
Constantino2011, Lin2015}). Participants were randomly assigned to either
condition (between-participants variation). 53 participants finished the
conflict negotiation condition and 41 finished the automatic condition. The
vignettes in both conditions focused on the process of handling possible
planning conflicts, but left the result open. Thus, we left open whether
a participant's request was granted in the final shift plan or not. We left the
final decision out, because we were primarily interested in the experience of
the process, not the result.

The ``conflict negotiation'' condition resembled the four interactive conflict
negotiation practices reported in~\citep{Uhde2021a} (see Appendix A for an
English translation of the full vignette texts). In the first vignette
(``rescheduling''), participants had requested a day off, but they were later
informed that a co-worker has also requested the same day off. They reacted by
attempting to reschedule their private appointment to a different day, where no
other worker has requested a day off, to increase the chance that it could be
granted. In the second vignette (``external resources''), participants had
requested a shift off for a private appointment they could not easily
reschedule. A co-worker requested the same shift off. They discussed with their
co-worker and tried to activate external resources so that both requests could
be granted (e.g., by asking healthcare workers from other groups to stand in or
by splitting the shift). The third vignette (``informal rules'') described an
expectable group conflict: Shift planning on Christmas\footnote{Christmas is an
important public holiday in the cultural context of our sample and many shift
workers want to spend it with their friends or families~\citep{Uhde2020a}.}.
Many workers wanted to get Christmas Eve off but not all requests could be
granted. Here, the group leader initiated an open discussion during a team
meeting based on an informal rule (i.e., ``whoever works on Christmas gets New
Year's Day off and vice versa''), again with an unknown result. Finally, the
fourth vignette (``unavoidable conflict'') described a conflict that could not
easily be resolved, and for which no informal rules existed. The employer
planned a company outing and healthcare workers from all wards of the
institution wanted to join. In addition, the outing overlapped with the holiday
season. Thus, the event could not be rescheduled, external resources were scarce
(e.g., workers from other groups), and there were no established, informal
rules. The healthcare workers searched for a solution together with a colleague
who also wanted to go, but it was unlikely that they found a solution that
allowed both of them to participate. In all four vignettes, the participants
were told that the algorithm made a decision if healthcare workers could not
find an alternative by themselves.

Participants in the ``automatic'' condition faced the same problems, but the
process did not involve the interactive conflict negotiation practices. Instead,
the decision was made by the system, which attempted to automatically find
a good solution (the result was not described in the vignettes). This is state
of the art in many automatic shift planning systems~\citep{Burke2004,
Constantino2011, Lin2015} and served as our control condition.

Following each vignette, we included several items to measure the dependent
variables. Subjective well-being was measured based on affect balance and need
fulfillment\footnote{We did not include the third measure of subjective
well-being, satisfaction with life~\citep{Diener1985}, because it is relatively
stable over time and effects of the few scenarios were unlikely.}. In addition,
we measured process fairness and workers' expected impact on team spirit. After
the fourth vignette was finished, we asked participants for their age, gender,
job title, and length of experience as a healthcare worker. Finally, at the end
of the questionnaire, we included an open field for comments about the study,
and a field to enter contact information for the raffle.

\begin{table*}[t]
  \caption{Summary statistics for all measures of the four vignettes and two
  resolution processes (n=94). m stands for the mean, sd stands
  for the standard deviation.}%
\label{tab:summary}
  \begin{tabular}{llSSSSSSSSSS}
\toprule

 process                   & scenario     & \multicolumn{2}{c}{immersiveness}
                           & \multicolumn{2}{c}{affect balance}
                           & \multicolumn{2}{c}{need fulfillment}
                           & \multicolumn{2}{c}{fairness}
                           & \multicolumn{2}{c}{team spirit} \\
                           &             & \multicolumn{1}{c}{m} & \multicolumn{1}{c}{sd}
                           &               \multicolumn{1}{c}{m} & \multicolumn{1}{c}{sd}
                           &               \multicolumn{1}{c}{m} & \multicolumn{1}{c}{sd}
                           &               \multicolumn{1}{c}{m} & \multicolumn{1}{c}{sd}
                           &               \multicolumn{1}{c}{m} & \multicolumn{1}{c}{sd} \\

\midrule

\multirow{5}{*}{negotiation} & rescheduling  & 6.36 & 1.03   & -0.62 & 2.93 & 3.01 & 1.02 & 4.85 & 1.66 & 4.42 & 2.10 \\
                           & resources       & 6.08 & 1.33   & -0.47 & 3.21 & 2.98 & 1.01 & 4.85 & 1.84 & 4.75 & 2.08 \\
                           & informal rules  & 6.36 & 1.20   &  0.43 & 3.00 & 2.99 & 1.07 & 5.51 & 1.59 & 5.13 & 1.72 \\
                           & unavoidable     & 5.32 & 1.89   & -1.60 & 3.03 & 2.51 & 1.03 & 4.49 & 2.22 & 4.11 & 1.95 \\
                           & overall         & 6.03 & 1.46   & -0.57 & 3.13 & 2.87 & 1.05 & 4.92 & 1.88 & 4.60 & 2.00 \\

\addlinespace

\multirow{5}{*}{automatic} & rescheduling    & 5.98 & 1.39   & -1.17 & 2.67 & 2.97 & 0.96 & 4.10 & 1.70 & 4.10 & 1.81 \\
                           & resources       & 6.07 & 1.26   & -2.59 & 2.17 & 2.62 & 0.92 & 3.59 & 1.55 & 3.34 & 1.59 \\
                           & informal rules  & 6.02 & 1.43   & -0.02 & 3.36 & 2.75 & 0.94 & 4.15 & 1.63 & 3.78 & 1.90 \\
                           & unavoidable     & 5.29 & 1.99   & -0.02 & 3.40 & 2.74 & 1.00 & 4.44 & 1.83 & 3.80 & 1.73 \\
                           & overall         & 5.84 & 1.57   & -1.00 & 3.11 & 2.77 & 0.96 & 4.07 & 1.70 & 3.76 & 1.78 \\

\bottomrule
\end{tabular}
\end{table*}

\subsection{Measures}

Summary statistics of all measures can be found in Table~\ref{tab:summary}.

\subsubsection{Immersiveness}

As is common practice in EVM~\citep{Aguinis2014, Atzmueller2010}, we measured
how well participants could immerse into the scenario with one 7-point scale
item ranging from 1 (``not at all'') to 7 (``very well''). Immersiveness of all
four vignettes was high across both conditions (means ranged between 5.29 and
6.36).

\subsubsection{Affect Balance}

We measured affect balance with two 7-point scales (positive and negative
affect) by asking how positive/negative participants felt during the described
situation. Both scales ranged from 1 (``not at all'') to 7 (``extremely''). We
calculated the affect balance as the difference between positive and negative
affect. Higher values represent a more positive affect balance.

\subsubsection{Psychological Need Fulfillment}

To measure psychological need fulfillment, we used the need scales from Sheldon
and colleagues~\citep{Sheldon2001} in a German
translation~\citep{Diefenbach2010}. We included the four needs that were
reported as relevant in shift planning in previous research~\citep{Uhde2021a}:
Autonomy, competence, popularity, and security. Each need was measured as the
average of three items and overall need fulfillment was measured as the average
of the four needs. An example item for autonomy was ``I felt free to do things
my own way.'' All items were measured with a 5-point scale from 1 (``not at
all'') to 5 (``extremely''). Internal consistencies of all needs in all four
scenarios were good (all between Cronbach's $\alpha=.76$ and $\alpha=.93$).

\subsubsection{Fairness}

As reported above, our study focused on fairness of the decision-making process,
not on the result. Accordingly, we measured the fairness of the process with one
7-point item adopted from~\citep{Uhde2020a}. The scale ranged from
1 (``unfair'') to 7 (``fair'').

\subsubsection{Expected Team Spirit}

Finally, we asked participants to estimate the long-term effect of the described
decision-making process on team coherence on a 7-point scale from
1 (``negative'') to 7 (``positive'').

\section{Results}

\subsection{Well-being}

\subsubsection{Affect Balance}

Our first hypothesis was that interactive conflict negotiation practices lead to
a more positive affect balance, compared with a fully automated process. We
tested this by running a 2 (resolution process) x 4 (scenario) ANOVA with affect
balance as measure. We did not find a main effect of the resolution process
($F(1,92)=0.98, p=.33$), but a significant interaction effect ($F(3,276)=7.53,
p<.01, \eta^2_{p}=.08$). A post-hoc analysis with Holm-adjusted $\alpha$ values
revealed a positive effect of the interactive ``external resources'' practice,
compared with the automated process ($t(92)=3.36, p<.01, d=.70$). There was no
effect for the other three interactive practices.

We further explored whether the effect of the ``external resources'' practice
was based on more positive affect or less negative affect, compared with the
automatic process. Both differences, for positive ($t(91.83)=3.63, p<.001,
d=.74$) and negative affect ($t(88.74)=3.29, p<.001, d=.66$), were significant.
Thus, the interactive practice led to more positive and less negative affect,
compared with the automatic process.

In sum, we could not confirm an overall positive effect for all interactive
conflict negotiation practices on the emotional experience of healthcare workers
(i.e., affect balance). However, we found a more specific, positive effect of
the ``external resources'' practice on the healthcare workers' emotional
experience, compared with an automated decision-making process. This effect was
based both on more positive affect and less negative affect.

\subsubsection{Need Fulfillment}

Our second hypothesis was that interactive conflict negotiation practices lead
to more need fulfillment, compared with the automatic process. We ran
a 2 (resolution process) x 4 (scenario) ANOVA with overall need fulfillment
as measure. Again, we found no main effect of the resolution process
($F(1,92)=0.85, p=.36$), but a significant interaction effect ($F(3,276)=7.02,
p<.01, \eta^2_{p}=.07$). A post-hoc analysis with Holm-adjusted $\alpha$ values
revealed a positive effect of the interactive ``external resources'' practice,
compared with the automatic process ($t(92)=2.33, p<.05$). The other practices
had no effect on need fulfillment.

We further explored the need fulfillment of the ``external resources'' practice
for the four needs separately, and found a significant difference in autonomy
($t(92)=2.95, p<.01, d=0.61$) and competence ($t(92)=2.48, p_{crit}=.025,
p=.015, d=0.52$). The differences in security and popularity were not
significant.

Taken together, we could not confirm an overall effect of the interactive
conflict negotiation practices on need fulfillment. However, we found a more
specific effect of the ``external resources'' practice on need fulfillment,
which was based on higher fulfillment of the autonomy and competence needs.

\subsection{Fairness}

Our third hypothesis was that interactive conflict negotiation practices are
perceived as fairer than a fully automated process. We ran a 2 (resolution
process) x 4 (scenario) ANOVA with fairness as measure and found a significant
main effect of the resolution process ($F(1,92)=10.31, p<.01, \eta^2_{p}=.10$).
We also found a significant interaction effect ($F(3,276)=4.04, p<.01,
\eta^2_{p}=.04$). A post-hoc analysis with Holm-adjusted $\alpha$ values
revealed a positive effect of the ``informal rules'' practice ($t(92)=3.67,
p<.01$) and the ``external resources'' practice ($t(92)=3.40, p_{crit}=.025,
p<.01$). The ``rescheduling'' and ``unavoidable conflict'' practices did not
reach significance after Holm-adjustment.

In sum, we could confirm a positive overall effect of the interactive conflict
negotiation practices on process fairness from the perspective of healthcare
workers, compared with automatic shift planning. This effect was particularly
based on positive effects of the ``external resources'' and ``informal rules''
practices.

\subsection{Team Spirit}

Our fourth hypothesis was that interactive conflict negotiation practices lead
to a more positive team spirit than a fully automated process. We ran
a 2 (resolution process) x 4 (scenario) ANOVA with team spirit as measure and
found a significant main effect of the resolution process ($F(1,92)=7.11, p<.01,
\eta^2_{p}=.07$). We also found a significant interaction effect
($F(3,276)=5.24, p<.01, \eta^2_{p}=.05$). A post-hoc analysis with Holm-adjusted
$\alpha$ values revealed a positive effect of the ``external resources''
practice ($t(92)=3.58, p<.01$) and the ``informal rules'' practice ($t(92)=3.43,
p_{crit}=.025, p<.01$). The effects of the ``rescheduling'' and ``unavoidable
conflict'' practices did not reach significance after Holm-adjustment.

Taken together, we could confirm a positive overall effect of the interactive
conflict negotiation practices on team spirit, compared with automatic shift
planning. This effect was specifically based on positive effects of the
``external resources'' and ``informal rules'' practices.

\begin{figure*}[t]
\centering
\includegraphics[width=\linewidth]{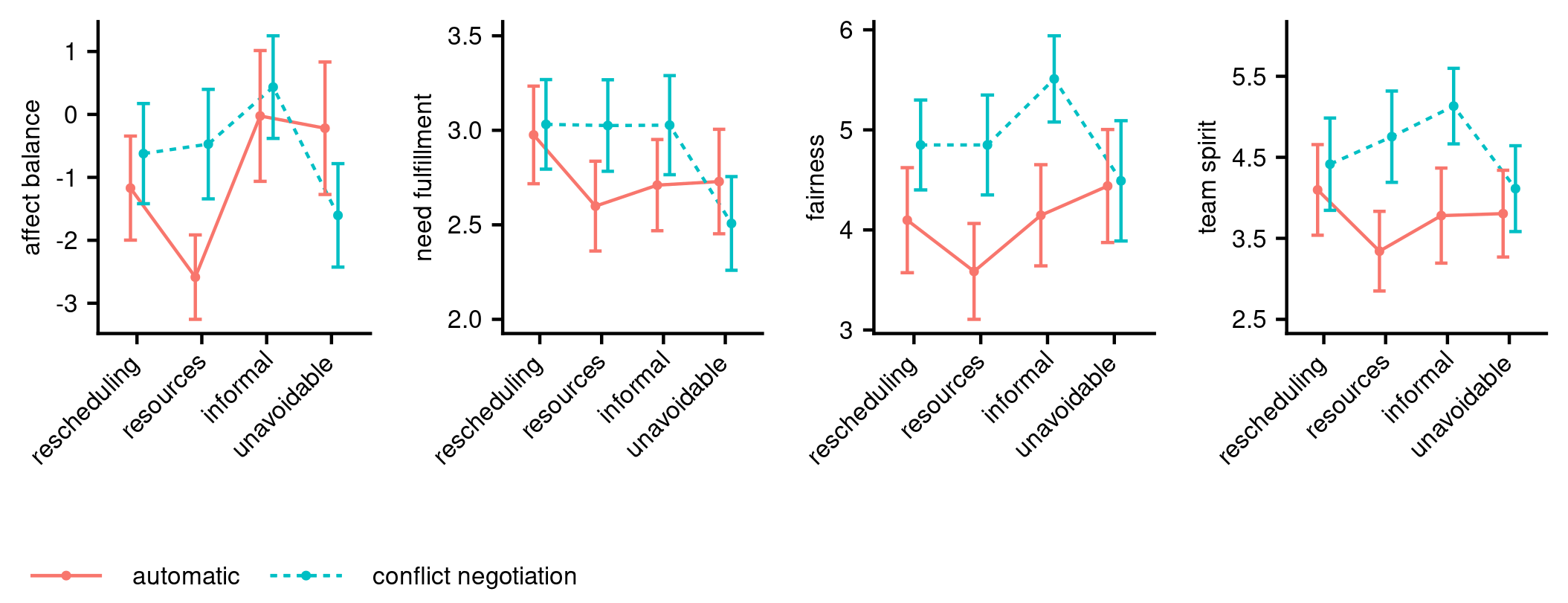}
\caption{Results for our four measures by resolution process and scenario. The
  error bars mark the 95\% confidence intervals.}
  \Description{Four line graphs, one for each measure (affect balance, need
  fulfillment, fairness, team spirit).}%
\label{fig:results}
\end{figure*}

\subsection{Open Comments}

In addition to the quantitative measures, 21 participants left a comment at the
end of our questionnaire (12 in the conflict negotiation condition, 9 in the
automatic condition). In this section, we present the comments that were
relevant to our research, organized by the different topics they addressed.

\subsubsection{Overall Evaluation of the Shift Planning Process}

The evaluation of the shift planning process seemed more positive in the
interactive conflict negotiation condition, compared with the automatic
condition. We received four positive comments on the process in the conflict
negotiation condition. One participant found it ``\textit{very interesting,
sounds good.}'', another one simply commented ``\textit{tip-top.}'' The other
two participants especially appreciated the long-term preferences feature we had
adopted from~\citep{Uhde2021a} (e.g., general preferences for morning shifts)
and briefly mentioned in the introduction. Notably, this feature was part of
both experimental conditions and did not play a role in the four scenarios.

In the automatic condition, one participant generally appreciated the interest
of academic researchers in shift planning. However, there were no other positive
comments on the shift planning process in this condition. One participant
expressed a negative opinion: ``\textit{The case-based questions can only be
answered in a negative way, because a system makes the decisions. Communication
among each other and conflict resolutions are missing.}'' The other comments
were not related to the automatic shift planning process.

\subsubsection{Conflict Negotiations}

Three further comments (two from the automatic condition) were related to
conflict negotiations more generally. One participant expressed that
``\textit{It is difficult, because you cannot please everyone.}'' Another one
also commented that ``\textit{It is difficult, especially with very emotional
decisions such as Christmas} [our scenario 3] \textit{which simply weighs more
than a company outing} [our scenario 4].'' The third comment highlighted the
importance of a positive team spirit and interpersonal relations:
``\textit{Depending on the colleague you work with, swapping a shift or so is
easier, or you are more considerate. Unfortunately.}''

\subsubsection{Structural Problems in Healthcare Shift Work}

Finally, three further comments (one from the automatic condition) were a bit
longer and described systemic problems in healthcare from the worker
perspective. These were not directly related to our research questions, but
relevant to shift planning more generally. One participant complained that
changes in the shift plan are not always transparent. This causes problems
especially for workers who only work a few hours each week, and who rely on the
good will of their colleagues to notify them about important changes to their
shift plans. Another participant described general problems that result from
shift work, such as health problems and difficulties to maintain a social life.
They suggested allowing shift workers to retire earlier, because of these
problems. Finally, the third participant argued that many difficulties in shift
planning result from a general staff shortage, as well as outdated, rigid
structures in the daily work organization.

\section{Discussion}


In this paper, we presented an experimental online study in which we compared
two different processes of handling conflicts between workers' requests in
computer-supported shift planning. The ``conflict negotiation'' process was
based on four interactive conflict negotiation practices that directly involved
workers in the decision-making process. First, healthcare workers tried to
\textit{reschedule} their own request to avoid a conflict with a colleague's
request. Second, they looked for \textit{external resources} (e.g., workers from
other wards, more flexible shift timings) to integrate both requests. Third,
healthcare workers used \textit{informal rules} to handle difficult but
expectable conflicts, for example around public holidays with many requests.
Fourth, workers attempted to find a good solution for all colleagues, even if
the conflict turns out to be \textit{unavoidable}. These interactive conflict
negotiation practices were drawn from realistic practices of healthcare workers
found in earlier work~\citep{Uhde2021a}. The other, ``automatic'' process was
based on fully automated conflict resolution of the same four problem scenarios,
as is currently state of the art in computer-supported shift planning systems
(e.g.,~\citep{Burke2004, Constantino2011, Lin2015, VandenBergh2012}).

Overall, we found that involving healthcare workers through interactive conflict
negotiation practices had positive effects on subjective process fairness and
team spirit, compared with the fully automated shift planning process. These
effects were especially pronounced for the ``external resources'' and ``informal
rules'' practices. We also found a more specific positive effect of the
``external resources'' practice on healthcare workers' subjective well-being
(i.e., emotional experience/affect balance and psychological need fulfillment).
In addition, our analysis of the (qualitative) open comments confirmed that
healthcare workers tended to appreciate the interactive process. In contrast,
a participant in the automatic condition complained that the system makes
decisions on their behalf, which takes away an opportunity to communicate and
resolve conflicts among colleagues. Notably, they framed the removal of conflict
resolution through automation as a problem, not as a desirable feature.




\subsection{Broader Impact}


Our findings further extend earlier work suggesting a positive effect of
interactive conflict negotiation practices on healthcare workers' well-being and
fairness. On the one hand, the semi-automated process indicates how we can
overcome challenges of worker overload, which can result from giving workers
full planning responsibility (e.g.,~\citep{Bailyn2007}). On the other hand, we
showed how a specific process can look like that gives workers direct and
meaningful control about the shifts they actually care about, and that does not
rely too much on external (even if human) decision-makers
(e.g.,~\citep{NabeNielsen2012, Roennberg2010}). A key difference to most work in
the automatic shift planning literature (e.g.,~\citep{Burke2004,
VandenBergh2012}) is that our process does not focus so much on ``resolving''
conflicts by making a decision for one worker and against another. Instead, the
practices are more focused on avoiding such dichotomous decisions, and trying to
account for both workers' needs while providing a functional shift plan. By
integrating both manual and automated components in the overall process, we get
the best of both worlds. We can benefit from the efficiency provided by
automation where appropriate, for example in planning the many shifts that are
not too relevant for individual workers or automating repeated preferences for
early shifts. In addition, we can benefit from more detailed context information
about the specific requests (e.g., is it somewhat flexible or not), and more
sensitive information about workers. These may not be available in automated
systems because of privacy concerns, technical limitations, and operational
costs, but may contribute to more satisfactory decision-making. Finally, we
found that some interactive conflict negotiation practices can not only help to
solve conflicts, but they can also have positive effects on the workers'
experiences, and thus contribute to a more fulfilling overall work experience.
In other words, our findings are encouraging for further research on hybrid
notions of computer-supported shift planning.

Beyond shift planning, our work also adds to the currently growing interest in
worker-centered design and the increasing focus on workers' values, needs, and
well-being in the development of workplace technologies
(e.g.,~\citep{CHIWORK2022, Karusala2021, Fox2020, Laloux2014, Laschke2020a,
Laschke2020b, Uhde2021a, Zhang2022}). Specifically, it offers a case of how work
practices that may seem ``mundane'', superfluous, or even annoying (such as
conflict resolution) can also be understood as valuable elements of the overall
work experience. Similar approaches in shift planning~\citep{Uhde2021a} and in
the communication between physicians with different
specialization~\citep{Laschke2020a} have already been presented, and our study
contributes measurable effects on workers' well-being. More broadly, this work
also relates to recent approaches to automation that do not focus on efficiency
alone, but primarily on promoting well-being and positive experiences of the
users (e.g.,~\citep{Klapperich2019b, Klapperich2020}).

In professional settings where people organize themselves in some way to work
together, an occasional conflict or disagreement might be unavoidable. In some
cases, technology plays a crucial role as the cause of conflicts, or it provides
the data workers can refer to when taking action to resolve the
conflicts~\citep{Clear2017, Dombrowski2017}. Such disagreements among workers or
between workers and management may not simply be ``resolved'' through automated
solutions that do not address underlying causes. For example, automated shift
planning decisions about who works on subjectively important days (e.g.,
Christmas in our study) are typically not desirable from a worker perspective.
Even if the final decision is made against their requests, workers want to have
a say in the decision-making process, as indicated in our study and in previous
work~\citep{Uhde2020a, Uhde2021a}. Thus, instead of circumventing the social
disagreement through automation, careful design of conflict resolution
processes, built on an understanding of the workers' needs and subjective
experiences, can lead to more positive outcomes (e.g., in terms of well-being,
fairness, and a positive team spirit).

\subsection{Design Implications}

\subsubsection{Optimize for Meaningful Involvement of Healthcare Workers in the Shift Planning Process}


Our study revealed specific, positive effects of worker involvement on
well-being, fairness, and team spirit. Notably, earlier work has also found
negative effects of other ways to involve workers in computer-supported shift
planning, for example on subjective control~\citep{Roennberg2010} and work
load~\citep{Bailyn2007}. From our perspective, it becomes increasingly clear
that the specific way this involvement is designed is essential for successful,
interactive shift planning. Simply reassigning planning tasks to the healthcare
workers for example by having them create full work schedules seems problematic,
because it primarily increases their workload. In addition, these tasks do not
seem to improve the healthcare workers' experiences, as long as they are not
meaningfully involved in the actual decision-making processes, or if these
processes do not consider the social dynamics in the team~\citep{Bailyn2007,
Roennberg2010}. Instead, shifting targeted control to the workers about the
specific work and free times they care about seems more helpful. In future shift
planning systems, the design of worker involvement should specifically be
optimized for meaningful worker control about these subjectively important
shifts.

This notion is further supported, considering that our participants also
appreciated certain forms of automation. Specifically, some participants
mentioned that the feature of automatically recurring long-term preferences such
as a weekly sports club (see also~\citep{Uhde2021a}) would be desirable. This is
a good example of an automation feature that adds value for the workers, does
not take away worker control for subjectively important requests, and reduces
their work load.

Finally, the most positive effects on workers' experiences resulted specifically
from the two practices that require direct interactions between workers. The
``external resources'' practice is based on a face-to-face negotiation between
two colleagues and led to higher subjective well-being, fairness, and team
spirit than an automated alternative. The ``informal rules'' practice is based
on a face-to-face negotiation in the team and led to higher fairness and team
spirit. Unlike the ``rescheduling'' practice, social interactions and
worker-based decision-making are inherent to these practices, and they cannot
simply be added as another feature to a shift planning system that is otherwise
supposed to be fully automated. Instead, they require an actual redesign of the
decision-making process that directly involves workers. Our findings indicate
that such a redesign would be beneficial from the worker perspective.

\subsubsection{Design Shift Planning to Improve Subjective Experiences of Workers}


Some earlier system proposals have already considered worker-oriented features
such as fairness~\citep{Constantino2011, Lin2015} or
well-being~\citep{Petrovic2019, Petrovic2020}. These features are typically
considered as ``objective'' criteria in the computer-supported shift planning
literature. In the fully automated shift planning paradigm, this ``objectivity''
is crucial, because it makes the criteria formalizable and accessible for
algorithmic processing. For example, Constantino and
colleagues~\citep{Constantino2011} suggested to (automatically) balance worker
requests that are integrated in the shift plan, to optimize fairness.

However, this ``objective'' approach has several problems. First, there is no
such thing as ``objective fairness'', because fairness can mean different things
in different contexts. It can be based on equality, equity (i.e., performance),
or need norms (and some derivates of the three~\citep{Deutsch1975}). These
different definitions are incompatible with each other~\citep{Kleinberg2017}.
Thus, some subjectivity is always involved, at least in the decision for an
underlying fairness norm, and automated solutions tend to be built on fairness
definitions that are incompatible with workers' actual fairness
experiences~\citep{Uhde2020a}. Second, fairness does not only concern the result
of a decision-making process, but also the process itself~\citep{Colquitt2001,
Colquitt2015b}. In shift planning, for example, involvement in the
decision-making process increases workers' subjective fairness, independent of
the result~\citep{Uhde2020a}. Finally, we think that fairness in shift planning
only makes sense in the first place if we talk about fairness from the
perspective of the affected healthcare workers. A solution that satisfies
a fairness definition of the algorithm developers does not help workers in their
everyday lives. To put it more drastically: We think that a completely unequal
solution that integrates five requests of one worker and only one request of
another worker can be preferable to one where both get three requests granted,
if that is what both workers think is the fairest solution. Such subjective
perspectives of healthcare workers should be considered more prominently in
future designs of computer-supported shift planning systems. At the end of the
day, the workers should be satisfied with the systems, rather than the designers
and their theoretical assumptions.

\subsubsection{Embed Conflict Negotiation Practices in the Broader Work Context}

We found positive effects of the interactive conflict negotiation practices, but
simply adding a worker control feature that enables them to negotiate may not be
sufficient to make the system work well in practice. Healthcare workers need to
stay motivated to participate over a longer time period if the system is
supposed to be sustainable~\citep{Bailyn2007}. Thus, socially embedding such
a shift planning system, guiding the workers through the process, and enabling
a cooperative culture, could be necessary for them to even engage in conflict
negotiations. On the one hand, the user interface could support this, for
example by indicating where conflicts may arise, where workers in other wards
may be available (because they have not submitted a request themselves), and by
allowing manual entries of flexible solutions. On the other hand, socially
embedding such a system and motivating healthcare workers to participate may
require good leadership skills of the group leader or other people in the
healthcare institution~\citep{Uhde2021a}.

One specific case where this could be especially important is the fourth
scenario of our study, where workers are faced with a seemingly unavoidable
conflict. Although the decision was delegated to the algorithm in our case, it
could be desirable if the group leader takes responsibility for such difficult
decisions, for a couple of reasons. First, these situations may be rather rare
(e.g., once in nine months in~\citep{Uhde2021a}), so efficiency is not a core
concern here. Second, human-made decisions have been found to increase workers'
experience of having voice in the decision-making process, compared with
algorithmic decisions~\citep{Schlicker2021a}. In other words, the mere fact that
a human makes the decision can signal that the workers' requests are taken
seriously (see also~\citep{Binns2020}). And third, communication between leaders
and workers plays a more generally important role for workers' fairness
perceptions in inclusive shift planning~\citep{Wynendaele2021}.



\subsection{Limitations and Further Research}


Although the inclusive shift planning system was well-received by the healthcare
workers in our study, there are some healthcare contexts in which such
negotiations that rely on interpersonal communication may be challenging in
practice. One such case is outpatient care, where colleagues may not meet each
other regularly. Of course, workers can use other forms of communication, but
the team dynamics may be more generally affected by such a work setting and
discourage pro-social behavior~\citep{Uhde2020b}. Thus, we think that shift
planning systems for such specific healthcare contexts need to be studied in
more detail in the future.

In addition, our study only covered practices of prospective shift planning.
Short-term changes of the shift plan, for example if some workers call in sick,
pose another challenge in shift work that we did not cover here. Creating new
shift plans and changing existing ones are often considered as separate
processes from a management perspective. But from the worker perspective, they
both affect their work and private time schedules. Thus, inclusive practices of
short-term changes are an additional important topic for future work.

As with any study, the method we chose (vignette-based experimental online
study) also comes with certain problems and advantages. Positive aspects include
that it allowed us to experimentally vary and control the exact conflict
negotiation practices in each condition, while preserving immersive and
realistic overall scenarios. But of course, an online study alone with
a specific sample from one cultural context is not sufficient. Future work
should include long-term studies using interactive prototypes that implement
these interactive conflict negotiation practices, and confirm or further specify
the positive effects we found.

Finally, shift planning itself needs to be understood within the broader
organization of healthcare in our societies. Of course, improving such central
processes can have positive effects. But the Covid-19 pandemic has revealed
severe systemic problems in the healthcare systems of many countries. Currently,
nursing is not an attractive job. Only few people want to enter the
profession~\citep{Hemsley1999}, and many healthcare workers retire early or
switch to different jobs~\citep{Chan2013}. Among the remaining healthcare
workers, there is a growing trend towards part-time instead of full-time
employment (e.g., in Germany~\citep{Pflegestatistik2001, Pflegestatistik2019}).
As our participants pointed out in the open comments, many problems of shift
planning are a consequence of the overall shortage of healthcare workers. This
leads to a situation where healthcare work becomes overwhelmingly exhausting to
many, and because of that one participant suggested allowing healthcare workers
to retire early. We think that fundamental changes are needed to make healthcare
work more attractive again, and to create more sustainable working conditions.
Improving shift planning processes is important in this regard, but to overcome
the structural problems, it needs to be complemented with fundamental
improvements to the healthcare system.

\section{Conclusion}

Our study contributes experimental findings about healthcare workers'
experiences during conflict negotiations in shift planning. Good,
worker-centered design of such central processes of work organization
successfully integrates the advantages of automation with meaningful worker
control. We see this work as one stepping stone that can help create better work
environments, and helps make future work in healthcare professions more
attractive again.


\begin{acks}

We would like to thank our participants and the healthcare managers who
forwarded our call for their support. We would also like to thank Esther Coban
for her support and early feedback from the perspective of a healthcare
professional on the design of the study. Finally, we would like to thank Mena
Mesenhöller for her support in designing and testing the study.

\end{acks}


\bibliographystyle{ACM-Reference-Format}
\bibliography{bibliography.bib}


\appendix

\clearpage
\section{Vignettes}

\begin{table}[h]
\scriptsize
  \caption{The four vignettes of the two resolution processes used in the study
  (translated from German).}%
\label{tab:vignettes}
  \begin{tabular}{cp{6.5cm}p{6.5cm}}
\toprule

    Vignette & Conflict Negotiation & Automatic \\
\midrule

    1 (rescheduling) &

    It is early in March and an old friend has recently contacted you. He will
    be in your area in the second week of April. You want to meet him and he has
    already said that the Friday would be good for him. Thus, you have already
    requested that Friday as a day off.

    During a shift you hear from a colleague, that she has also requested the
    same day off. Her grandfather will turn 90 on that day and she would like to
    go to the family party. From experience you know that requests can not
    always be integrated into the schedule, for example if several colleagues
    request the same day off.

    You send a short message to your friend, asking if Wednesday or Thursday
    would also be good for him. On these days, nobody else has requested a day
    off yet. If it works, you can change your request. Otherwise you cannot say
    for certain, whether you will get the day off or not. The system normally
    handles such conflicts well and it will automatically try to include as many
    requests as possible. &

    It is early in March and an old friend has recently contacted you. He will
    be in your area in the second week of April. You want to meet him and he has
    already said that the Friday would be good for him. Thus, you have already
    requested that Friday as a day off.

    During a shift you learn that a colleague has also requested the same day
    off. Her grandfather will turn 90 on that day and she would like to go to
    the family party. From experience you know that requests can not always be
    integrated into the schedule, for example if several colleagues request the
    same day off.

    You send a short message to your friend and tell him that you cannot say for
    certain, whether you will get the day off or not. The system normally
    handles such conflicts well and it will automatically try to include as many
    requests as possible. \\

    \addlinespace

    2 (resources) &

    You have made a doctor's appointment for early May. With this doctor it is
    difficult to get appointments. If you cancel an appointment, it often takes
    weeks to get a new one. You have requested this appointment as a free shift
    on the tablet.

    During a break a colleague tells you that he has requested the same shift
    off.  He is a single father and has to fetch his child from the daycare
    center earlier than usual on that day. You know that other colleagues are on
    vacation or have a professional training on that day, so that probably one
    of you two has to work.

    You try to find a solution together. Maybe a healthcare worker from
    a different group can help out? Maybe you can split the shift somehow so
    that he works the first half and you the second? If you find a solution, you
    can directly enter it in the system and it will be integrated in the plan.
    Otherwise the system tries to automatically find a good solution, even in
    such tense situations. However, not all requests can be granted in every
    case. &

    You have made a doctor's appointment for early May. With this doctor it is
    difficult to get appointments. If you cancel an appointment, it often takes
    weeks to get a new one. You have requested this appointment as a free shift
    on the tablet.

    During a break you learn that a colleague has requested the same shift off.
    He is a single father and has to fetch his child from the daycare center
    earlier than usual on that day. You know that other colleagues are on
    vacation or have a professional training on that day, so that probably one
    of you two has to work.

    Even in such tense situations the system tries to automatically find a good
    solution. However, not all requests can be granted in every case.

    \\

    \addlinespace

    3 (informal rule) &

    The shift plans for December will have to be made soon. Normally most
    colleagues either want Christmas or New Year's Day off and they enter their
    requests. Just like many other colleagues, you would like to have Christmas
    off this year. Based on experience, not all requests can always be granted.

    In your group you have come up with an informal rule: Whoever works on
    Christmas gets New Year's Eve off and vice versa. During a team meeting your
    group leader brings up the topic: Who works when? Together you try to find
    a solution, so that everyone has either Christmas or New Year's Day off. If
    you find such a solution, it will directly be integrated into the schedule.
    If that does not work, the system will try to automatically find a good
    solution, so that as many requests as possible can be granted. &

    The shift plans for December will have to be made soon. Normally most
    colleagues either want Christmas or New Year's Day off and they enter their
    requests. Just like many other colleagues, you would like to have Christmas
    off this year. Based on experience, not all requests can always be granted.

    As soon as everyone has entered their requests, the system will try to
    automatically find a good solution, so that as many requests as possible can
    be granted. \\

    \addlinespace

    4 (unavoidable) &

    Your employer has planned a company outing. Healthcare workers from all
    groups want to participate and so do you. Of course the company needs to
    keep running. Unfortunately the date is also during the holiday season, so
    that some colleagues are not there. The staff situation is already tense.

    A colleague from your group tells you that she also wants to go. But it is
    unlikely that both of you can go. You discuss all the options, but it is
    difficult. The date cannot be changed and you cannot ask anyone from other
    groups, because they are also short on staff. The last company outing
    happened a few years ago and your colleague has not work in your company at
    the time. There does not seem to be a solution for both of you to go. In the
    end, you simply both submit a request on the tablet and let the system
    decide who should go. &

    Your employer has planned a company outing. Healthcare workers from all
    groups want to participate and so do you. Of course the company needs to
    keep running. Unfortunately the date is also during the holiday season, so
    that some colleagues are not there. The staff situation is already tense.

    You learn that a colleague from your group also wants to go. But it is
    unlikely that both of you can go. You submit your request on the tablet and
    let the system decide who should go. \\

\bottomrule
\end{tabular}
\end{table}

\end{document}